\documentclass[fleqn,usenatbib]{mnras}

\usepackage{newtxtext,newtxmath}

\usepackage[T1]{fontenc}

\DeclareRobustCommand{\VAN}[3]{#2}
\let\VANthebibliography\thebibliography
\def\thebibliography{\DeclareRobustCommand{\VAN}[3]{##3}\VANthebibliography}

\usepackage{graphicx}	
\usepackage{amsmath}	

\usepackage{tikz}

\newcommand\pU{\mathcal{U}}

\title[Direct inference from 3D Ly$\alpha$ correlations]{Direct cosmological inference from three-dimensional correlations of the Lyman-$\alpha$ forest}

\author[Francesca Gerardi et al.]{
Francesca Gerardi,$^{1}$\thanks{E-mail: francesca.gerardi.19@ucl.ac.uk}
Andrei Cuceu,$^{2,3,4}$\thanks{E-mail: cuceu.1@osu.edu}
Andreu Font-Ribera,$^{5,1}$
Benjamin Joachimi$^{1}$
and Pablo Lemos$^{6,1}$
\\
$^{1}$ Department of Physics \& Astronomy, University College London, Gower Street, London WC1E 6BT, UK\\
$^{2}$ Center for Cosmology and Astro-Particle Physics, The Ohio State University, Columbus, Ohio 43210, USA\\
$^{3}$ Department of Astronomy, The Ohio State University, Columbus, Ohio 43210, USA\\
$^{4}$ Department of Physics, The Ohio State University, Columbus, Ohio 43210, USA\\
$^{5}$ Institut de Física d’Altes Energies, The Barcelona Institute of Science and Technology, Campus UAB, 08193 Bellaterra (Barcelona), Spain\\
$^{6}$ Department of Physics and Astronomy, University of Sussex, Sussex House, Falmer, Brighton, BN1 9RH, UK
}

\date{Accepted XXX. Received YYY; in original form ZZZ}

\pubyear{2022}

\begin{document}
\label{firstpage}
\pagerange{\pageref{firstpage}--\pageref{lastpage}}
\maketitle

\begin{abstract}
When performing cosmological inference, standard analyses of the Lyman-$\alpha$ (Ly$\alpha$) three-dimensional correlation functions only consider the information carried by the distinct peak produced by baryon acoustic oscillations (BAO). In this work, we address whether this compression is sufficient to capture all the relevant cosmological information carried by these functions. We do this by performing a direct fit to the full shape, including all physical scales without compression, of synthetic Ly$\alpha$ auto-correlation functions and cross-correlations with quasars at effective redshift $z_{\rm{eff}}=2.3$, assuming a DESI-like survey, and providing a comparison to the classic method applied to the same dataset.
Our approach leads to a $3.5\%$ constraint on the matter density $\Omega_{\rm{M}}$, which is about three to four times better than what BAO alone can probe. The growth term $f \sigma_{8} (z_{\rm{eff}})$ is constrained to the $10\%$ level, and the spectral index $n_{\rm{s}}$ to $\sim 3-4\%$. 
We demonstrate that the extra information resulting from our `direct fit' approach, except for the $n_{\rm{s}}$ constraint, can be traced back to the Alcock-Paczyński effect and redshift space distortion information.
\end{abstract}

\begin{keywords}
cosmological parameters -- large-scale structure of universe --  methods: data analysis
\end{keywords}



\section{INTRODUCTION}

Over the last couple of decades, after the discovery of the accelerated expansion of the Universe \citep{Riess_1998, Perlmutter_1999}, cosmology has focused on investigating the properties of dark energy. Among the multiple probes used to place a contraint on the parameters of the current $\Lambda$CDM model, there is the baryon acoustic oscillation (BAO) scale over different tracers\footnote{For a BAO distance ladder plot \url{https://www.sdss.org/science/cosmology-results-from-eboss/}.} of the matter density field \citep{SDSS:2005xqv, 2dFGRS:2005yhx}. Measurements of this standard ruler over a range of redshifts place a constraint on the expansion history \citep{2003ApJ...598..720S}.

Complementary to low-redshift galaxies ($z \lesssim 1$),  the Lyman-$\alpha$ (Ly$\alpha$) forest is a tracer of the intergalactic medium (IGM) that probes the cosmic expansion via BAO at higher redshifts, as first proposed by \citet{PhysRevD.76.063009}. The Ly$\alpha$ forest is a sequence of absorption lines in high-redshift quasar (QSO) spectra, caused by the neutral hydrogen distributed along the line of sight, between the quasar and the observer. The first  BAO detection from the Ly$\alpha$ auto-correlation function and from its cross-correlation with QSOs was in the Baryon Oscillation Spectroscopic Survey (BOSS) DR9 data \citep{Busca_Delubac_Rich_Bailey_Font-Ribera_Kirkby_Le_Goff_Pieri_Slosar_Aubourg_et_al._2013, Slosar:2013fi, Kirkby:2013fh} and DR11 data \citep{BOSS:2013igd}, respectively. 

BAO produce a distinct feature in the correlation functions, which we wish to measure and use to probe cosmology, in a robust and model-independent way. When performing cosmological inference, a standard method, as applied in BOSS and eBOSS \citep{Bourboux_Rich_Font-Ribera_Agathe_Farr_Etourneau_Goff_Cuceu_Balland_Bautista_et_al._2020} 
analyses of the Ly$\alpha$ three-dimensional correlation functions, relies on splitting them into a \textit{peak} and a \textit{smooth} component and only considers the information carried by the BAO feature.

Recently, \citet{Cuceu_Font-Ribera_Joachimi_Nadathur_2021} (C21 hereafter) demonstrated that further cosmological information can be obtained from the broadband component using the Alcock-Paczyński (AP) effect \citep{1979Natur.281..358A}. When computing the 3D correlation functions from observations, as a standard approach, we change from angular and redshift separations to comoving coordinates, based on the assumption of a fiducial cosmological model. In particular, if the latter differs from the true underlying cosmology, then the AP effect will appear as an apparent anisotropy in the correlation functions. Another source of anisotropy is redshift space distortions (RSD), which are induced by peculiar velocities and hence carry extra information. However, measuring redshift space distortions for the Ly$\alpha$ auto-correlation alone is not informative about the growth rate of structure because of its degeneracy with an unknown velocity divergence bias \citep{Seljak_2012}. For this reason, C21 jointly employed the Ly$\alpha$ auto- and cross-correlation with quasars to explore the potential of measuring the linear growth of structure.

All physical scales of the 3D Ly$\alpha$ correlation functions, beyond the BAO peak, carry information about the underlying cosmology. Throughout this work we will refer to the sum of all of these scales, with no compression, as the full shape of these functions. Both the Ly$\alpha \times$Ly$\alpha$ auto- and Ly$\alpha \times$QSO cross-correlation functions can be directly used to perform cosmological inference. The work of C21 motivates a further investigation, assessing whether or not the compressed analysis based on BAO, AP and RSD successfully captures all cosmological information from the correlation functions of interest.  

The same point is relevant also in the field of galaxy clustering, where the compressed standard approach extracts cosmological information from BAO, AP and RSD \citep{2017MNRAS.470.2617A, 2021PhRvD.103h3533A}. Over the past few years, advancements in perturbation theory computations boosted the interest in fitting the observed two-point statistics and directly inferring cosmological parameters without compression \citep{troester20,Ivanov_2020,d_Amico_2020}. In particular, \citet{Brieden_2021} carried out such an investigation, identifying from where in the data vector additional information originates and extending the classic approach by introducing an extra physical parameter.

In this work, we aim to address whether or not compression is a suitable approach in the field of Ly$\alpha$ forest cosmology, by performing a direct fit to the full shape of synthetic Ly$\alpha \times$Ly$\alpha$ auto- and Ly$\alpha \times$QSO cross-correlation functions, and by subsequently comparing the constraints with those obtained using the standard approach. 
In real data, Ly$\alpha$ correlations have contaminants that affect the amount of cosmological information extractable from a given scale.
We discuss here an optimistic scenario without contaminants, and leave for future work a detailed study of their impact on the cosmological constraints.

The paper is structured as follows. We start in Sect. \ref{sect:lya_method} by outlining the methodology and explaining the inference framework we use. We proceed to apply this method to noiseless synthetic correlation functions and present the forecasts in Sect. \ref{sect:lya_forecasts}. We finally compare our main results to the compressed analyses in Sect. \ref{sect:lya_discussion} and draw our conclusions in Sect. \ref{sect:lya_conclusions}.

\section{METHOD}\label{sect:lya_method}

We use the full shape of the 3D Ly$\alpha$ correlation functions to directly infer cosmological parameters, without the usual compression methods. Based on the modelling of C21 and with the aid of \textsc{camb} \citep{Lewis_2000, Howlett_2012}, we construct a likelihood for the \textsc{cobaya} framework \citep{Torrado_2021, 2019ascl.soft10019T}, and we perform Markov chain Monte Carlo (MCMC) sampling \citep{PhysRevD.66.103511, https://doi.org/10.48550/arxiv.math/0502099, PhysRevD.87.103529} using the Ly$\alpha \times$Ly$\alpha$ auto- and Ly$\alpha \times$QSO cross-correlations.

In this section we outline the key features of the method. In Sect. \ref{subsect:lya_mocks} we describe how the synthetic correlation data was generated, and focus on the modelling in Sect. \ref{subsect:lya_method_model}. In particular, throughout the analysis, we use the code \textsc{vega}\footnote{\url{https://github.com/andreicuceu/vega}} \citep{Cuceu_2020}, which is based on the code used in eBOSS DR16 \citep{Bourboux_Rich_Font-Ribera_Agathe_Farr_Etourneau_Goff_Cuceu_Balland_Bautista_et_al._2020} for fitting and modelling the Ly$\alpha$ correlation functions. We finally motivate the choice of the sampled parameter space and describe the likelihood in Sect. \ref{subsect:lya_method_likelihood}. 

\begin{table*}
\centering
\begin{tabular}{ l c c c}
\hline
Parameter &  Fiducial & Prior & 68\% limits \\
\hline
$H_0 [\rm{km/ (s \times Mpc)}]$ & $67.31$ & $\pU(40,100)$ & $67.69^{+5.5}_{-3.16}$\\
$\Omega_{\rm{M}}$ & $0.3144$ & $\pU(0.01,0.99)$ & $0.318 \pm 0.011$ \\
$\Omega_{\rm{B}} h^2$ & $0.02222$ & $\pU(0.01,0.05)$ & $0.0229^{+0.0064}_{-0.0038}$ \\
$A_{\rm{s}}$ & $ 2.196 \cdot 10^{-9}$ & $\pU_{\rm{log(10^{10} A_{\rm{s}})}}(0.5,6)$ & $\left(\,2.06^{+0.42}_{-0.46}\,\right)\cdot 10^{-9}$\\ 
$n_{\rm{s}}$ & $0.9655$  & $\pU(0.8,1.2)$ & $0.958^{+0.025}_{-0.035}$ \\ \hline 
$b_{\rm{Ly}\alpha}$ & $-0.117  $  & $\pU_{\rm{log(-b_{Ly\alpha})}}(-2,0)$ & $-0.111^{+0.011}_{-0.012} $\\
$\beta_{\rm{Ly}\alpha}$ & $1.67$  & $\pU(0,5)$ &  $1.67 \pm 0.03$\\
$b_{\rm{QSO}}$ & $3.8$  &  $\pU_{\rm{log(b_{\rm{QSO}})}}(-2,1.3)$ & $ 3.61^{+0.47}_{-0.32}$\\
$\sigma_{\rm{v}} ({\rm Mpc}/h)$ & $6.86 $  &  $\pU(0,15)$ & $6.75^{+0.64}_{-0.55}$\\
\end{tabular}
\caption{Full set of sampled parameters, alongside with the fiducial values used to compute the synthetic correlations and the uniform ($\pU$) priors adopted for the sampling procedure. When sampling in logarithmic space we add a `log' subscript to $\pU$. In the last column, we provide the one-dimensional marginals (68\% c.l.) for all the parameters sampled, where for any asymmetric posteriors we report the posterior maximum with lower and upper 68\% limits.}
\label{tab:full_table}
\end{table*}

\subsection{Synthetic data vector and covariance} \label{subsect:lya_mocks}

In this work, we focus on idealised 3D Ly$\alpha$ synthetic correlations in flat $\Lambda$CDM, without contaminants. However, we do include the distortion due to quasar continuum fitting. The latter filters out information and `distorts' the true correlation function \citep{Bautista:2017zgn,dMdBourboux_2017}. Our synthetic data was generated using the framework of C21, and is given by an uncontaminated model based on the best fit of eBOSS DR16 (see Tab.~\ref{tab:full_table}). We did not add noise to the data vector, as we are only interested in forecasting. We used covariance matrices based on DESI mocks similar to those used in \citet{https://doi.org/10.48550/arxiv.2205.06648}. These mocks were created with the CoLoRe \citep{Ram_rez_P_rez_2022} and LyaCoLoRe \citep{Farr_2020} packages, covering 14000 sq. degrees with a target density of $\sim$ 50 QSOs / sq. degree \citep{https://doi.org/10.48550/arxiv.1611.00036}. The covariance was computed using the community package \texttt{picca} \citep{Bourboux_Rich_Font-Ribera_Agathe_Farr_Etourneau_Goff_Cuceu_Balland_Bautista_et_al._2020}. In this analysis, we limit ourselves to linear scales, assuming $r_{\rm{min}} = 30 \rm{h^{-1}Mpc}$, up to $r_{\rm{max}} = 180 \rm{h^{-1}Mpc}$. The effective redshift of the correlation functions is $z_{\rm{eff}} = 2.3$. 

\subsection{Modelling}\label{subsect:lya_method_model}

To infer cosmology from these synthetic correlations, we first need a theory to model the data given any cosmology $\boldsymbol{p_{\rm{C}}}$. The theoretical 3D Ly$\alpha$ correlation functions are computed from the isotropic matter power spectrum $P(k)$ and then compared against data to evaluate the likelihood.

When modelling and fitting Ly$\alpha$ correlations, we must match the coordinate grid for the theoretical correlation $\boldsymbol{\xi}$ with the grid of the data. When measuring the 3D Ly$\alpha$ correlation functions from observations, we change from angular $\Delta \theta$ and redshift $\Delta z$ separations to a set of comoving coordinates $(r_{\parallel},r_{\perp})$, respectively defined along and across the line of sight. This is motivated by the fact that both the radial comoving distance $D_{\rm{C}} (z) = c \int^{z}_{0} dz/H(z)$ and the comoving angular diameter distance $D_{\rm{M}}(z)$ are redshift dependent, where $c$ is the speed of light and $H(z)$ the Hubble parameter. For this reason, we wish to refer instead to a set of comoving coordinates. Given two locations at redshift $z_i$ and $z_j$ separated by an angle $\Delta \theta$, these are defined as

\begin{align}
    r_{\parallel} & = [D_{\rm{C,fid}}(z_i)-D_{\rm{C,fid}}(z_j)] \; \textrm{cos} \; \frac{\Delta \theta}{2} \; ; \\
    r_{\perp} & = [D_{\rm{M,fid}}(z_i)+D_{\rm{M,fid}}(z_j)] \; \textrm{sin} \; \frac{\Delta \theta}{2} \; ,
\end{align}
where both $D_{\rm{C}}$ and $D_{\rm{M}}$ are computed using an assumed fiducial cosmology ($\rm{fid}$ subscript). In our case, the fiducial cosmology coincides with the cosmological model that was used to generate the data vector. Given that the sampled cosmology that generated the theoretical $\boldsymbol{\xi}$ can be different from the fiducial one, we need to match coordinate grids of data and $\boldsymbol{\xi}$ by rescaling at each sampling step the coordinates of the correlation via
\begin{align}
    q_{\parallel} & = D_{\rm{H}}(z_{\rm{eff}})/D_{\rm{H}}^{\rm{fid}}(z_{\rm{eff}}) \; ; \\
    q_{\perp} & = D_{\rm{M}}(z_{\rm{eff}})/D_{\rm{M}}^{\rm{fid}}(z_{\rm{eff}}) \; ,
\end{align}
where $D_{\rm{H}}(z)=c/H(z)$, such that $r_{\parallel, \perp}^{'} = q_{\parallel, \perp} r_{\parallel, \perp}$.

In modelling the Ly$\alpha$ correlation functions of interest we follow Eq.~(27) of \citet{Bourboux_Rich_Font-Ribera_Agathe_Farr_Etourneau_Goff_Cuceu_Balland_Bautista_et_al._2020}, adopting the prescriptions of C21.  
For any cosmology $\boldsymbol{p_{\rm{C}}}$, the power spectra of the tracers are computed from the isotropic linear matter power spectrum $P(k,z)$ as

\begin{align}
    P_{\rm{Ly}\alpha}(k,\mu_k,z) = & b_{\rm{Ly}\alpha}^{2} \left( 1+\beta_{\rm{Ly}\alpha}\mu_k^2 \right)^{2} F_{\rm{nl, Ly}\alpha}^2(k,\mu_k)P(k,z) \label{eqn:lyalyapower} \; ; \\
    P_{\times}(k,\mu_k,z) = & b_{\rm{Ly}\alpha} \left( 1+\beta_{\rm{Ly}\alpha}\mu_k^2 \right) \nonumber\\
    &  \times \left( b_{\rm{QSO}}+f(z)\mu_k^2 \right) F_{\rm{nl,QSO}}(k_{\parallel})P(k,z) \; , \label{eqn:lyaqsopower}
\end{align}
with $f(z) = \frac{d\;\rm{ln}\;D}{d\;\rm{ln}\;a}$ being the logarithmic growth rate and $\mu_k = k_{\parallel}/k$, where $k$ and $k_{\parallel}$ are the modulus of the wave vector and its projection along the line of sight respectively. Focusing first on the Ly$\alpha \times$Ly$\alpha$ power spectrum in Eq.~(\ref{eqn:lyalyapower}), we identify the Ly$\alpha$ forest linear bias, $b_{\rm{Ly}\alpha}$, and its RSD term, $\beta_{\rm{Ly}\alpha} = \frac{b_{\rm{\eta, Ly}\alpha} f(z)}{b_{\rm{Ly}\alpha}}$, where $b_{\rm{\eta, Ly}\alpha}$ is an extra unknown bias, the velocity divergence bias. The choice of using $\beta_{\rm{Ly}\alpha}$ in the RSD term comes from the fact that $b_{\rm{\eta, Ly}\alpha}$ is fully degenerate with the growth rate $f(z)$. We treat both $b_{\rm{Ly}\alpha}$ and $\beta_{\rm{Ly}\alpha}$ as nuisance parameters and marginalize over them. The $F_{ \rm{nl,Ly}\alpha}$ term encodes the non-linear corrections according to the model of \citet{Arinyo-i-Prats:2015vqa}. The parameters involved in this model are kept constant for simplicity, but in principle they should also be varied. However, this should not have a major impact in our analysis, as we are restricting the analysis to linear scales ($r_{\rm{min}} = 30 \rm{h^{-1}Mpc}$). Moving on to the  Ly$\alpha \times$QSO power spectrum $P_{\times}$ in Eq.~(\ref{eqn:lyaqsopower}), $b_{\rm{QSO}}$ is the quasar linear bias, another nuisance parameter. In contrast to the Ly$\alpha$ RSD term, the QSO RSD term is instead simply $f(z)$ as by definition $b_{\rm{\eta, QSO}} = 1$. Following \citet{Bourboux_Rich_Font-Ribera_Agathe_Farr_Etourneau_Goff_Cuceu_Balland_Bautista_et_al._2020}, we model the impact of redshift errors and non-linear peculiar velocities of quasars with a damping term $F_{\rm{nl,QSO}}(k_{\parallel})$. We use a Lorentzian function 

\begin{equation}
    F_{\rm{nl, QSO}}(k_{\parallel}) = \sqrt{\left[ 1+ \left( k_{\parallel}\sigma_{\rm{v}} \right)^2 \right]^{-1}}   \; ,
\end{equation}
with a free parameter $\sigma_{\rm{v}}$, which represents the velocity dispersion and is an extra nuisance parameter.

At each sampling step then, theoretical correlation functions $\boldsymbol{\xi}$ are computed in \textsc{vega}. At its core, \textsc{vega} decomposes the power spectrum into multipoles, transforms them into correlation function multipoles using the FFTLog algorithm \citep{10.1046/j.1365-8711.2000.03071.x} and finally reconstructs the two-dimensional correlation function.

\subsection{Parameter space and likelihood}\label{subsect:lya_method_likelihood}

The BAO feature is able to constrain $\alpha_{\parallel} = D_{\rm{H}}(z)r_{\rm{d}}^{\rm{fid}}/D_{\rm{H}}^{\rm{fid}}(z)r_{\rm{d}}$ along the line of sight and $\alpha_{\perp} = D_{\rm{M}}(z)r_{\rm{d}}^{\rm{fid}}/D_{\rm{M}}^{\rm{fid}}r_{\rm{d}}$ in the transverse direction, where $r_{\rm{d}}$ is the sound horizon at the drag epoch. Since in flat $\Lambda$CDM, at low redshifts, the Hubble parameter $H(z)$ can be expressed as a function of $H_0$ and ${\Omega_{\rm{M}}}$, $\alpha_{\parallel}$ and $\alpha_{\perp}$ will ultimately place a constraint on $ \{ H_0r_{\rm{d}}, \Omega_{\rm{M}} \}$. Additionally, $r_{\rm{d}}$ can be numerically approximated as a function of $\Omega_{\nu}h^2$, $\Omega_{\rm{M}}h^2$ and $\Omega_{\rm{B}}h^2$ \citep{PhysRevD.92.123516}, which are the neutrino, matter and baryon densities respectively, all evaluated at redshift $z =0$ by definition. This further motivates the choice of sampling $\{H_0, \Omega_{\rm{M}}, \Omega_{\rm{B}} h^2\}$, where $\Omega_{\nu}h^2$ is constant for a given choice of the neutrino mass. Extra information on ${\Omega_{\rm{M}}}$ also comes from the AP effect \citep{1979Natur.281..358A}
\begin{equation}\label{eqn:AP_effect}
    F_{\rm{AP}} = \dfrac{a_{\perp}}{a_{\parallel}} = \dfrac{D_{\rm{M}}(z_{\rm{eff}})/D_{\rm{M}}^{\rm{fid}}(z_{\rm{eff}})}{D_{\rm{H}}(z_{\rm{eff}})/D_{\rm{H}}^{\rm{fid}}(z_{\rm{eff}})} = \dfrac{[D_{\rm{M}}(z_{\rm{eff}})H(z_{\rm{eff}})]}{[D_{\rm{M}}(z_{\rm{eff}})H(z_{\rm{eff}})]_{\rm{fid}}},
\end{equation}
which is an apparent anisotropy present if the sampled cosmology differs from the fiducial one. 
For the same assumptions as before, the AP parameter will only be a function of $\Omega_{\rm{M}}$. As we fit the full shape of the correlation functions directly and the amplitude of primordial fluctuations $A_{\rm{s}}$ and their spectral index $n_{\rm{s}}$ affect the functional form of $\boldsymbol{\xi}$, we will sample the full set of parameters $\boldsymbol{p_{\rm{C}}}=\{H_0, \Omega_{\rm{M}}, \Omega_{\rm{B}} h^2,A_{\rm{s}},n_{\rm{s}}\}$. On the other hand, $ \{b_{\rm{Ly}\alpha}, b_{\rm{QSO}}, \beta_{\rm{Ly}\alpha}, \sigma_{\rm{v}}\}$ are treated as nuisance parameters $\boldsymbol{p_{\rm{A}}}$ to marginalize over. 

For all these parameters we choose uniform priors, which are listed in Tab.~\ref{tab:full_table}. As is common, $A_{\rm s}$ is sampled in logarithmic space, and we made the choice of doing the same with the two linear biases because they are degenerate with $A_{\rm s}$ and span over several orders of magnitude.

In this work, we assume a Gaussian likelihood, which is also computed using \textsc{vega}. A likelihood evaluation, via \textsc{camb}, first computes the comoving distances, to calculate $q_{\parallel}$ and $q_{\perp}$, and then the isotropic linear matter power spectrum, along with $r_{\rm{d}}$, the growth rate at $z_{\rm{eff}}$ and $f \sigma_{8} (z_{\rm{eff}})$. Then, \textsc{vega} computes the correlation functions, based on the modelling description in Sect. \ref{subsect:lya_method_model}, and the $\chi^2$ value.

\begin{figure*}
    \centering
    \includegraphics[scale=0.335]{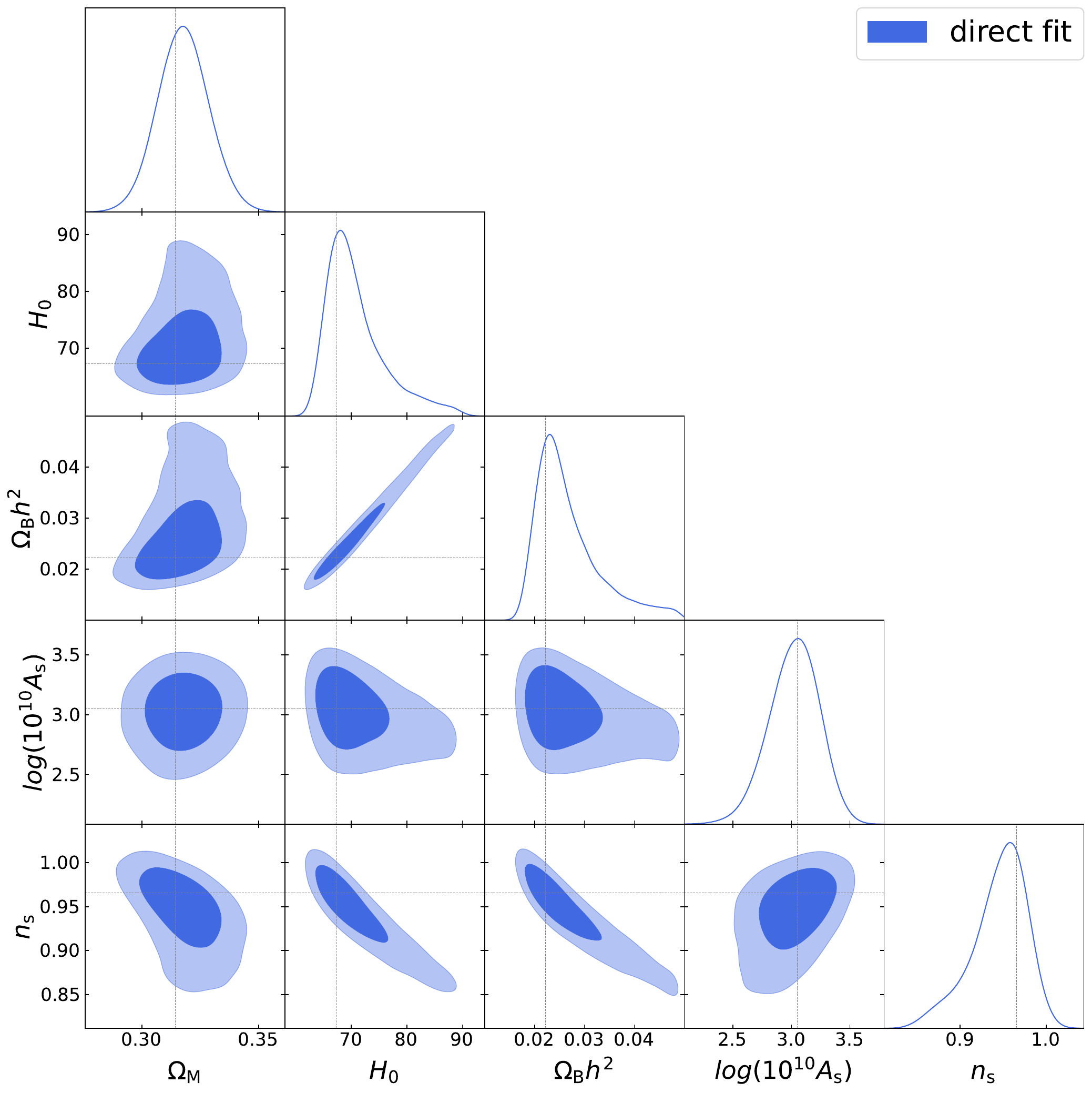}
    \caption{Triangle plot of the cosmological parameters of interest $\{ H_0, \Omega_{\rm{M}}, \Omega_{\rm{B}} h^2, A_{\rm{s}}, n_{\rm{s}} \}$, marginalizing over the nuisance parameters $\boldsymbol{p_{\rm{A}}}$. The blue contours refer to the results obtained performing the inference using the method outlined in Sect. \ref{sect:lya_method}, which we denote as `direct fit'. The grey dashed lines mark the fiducial values used to generate data. }
    \label{fig:growth_cosmo}
\end{figure*}

\begin{figure}
    \centering
    \includegraphics[scale=0.24]{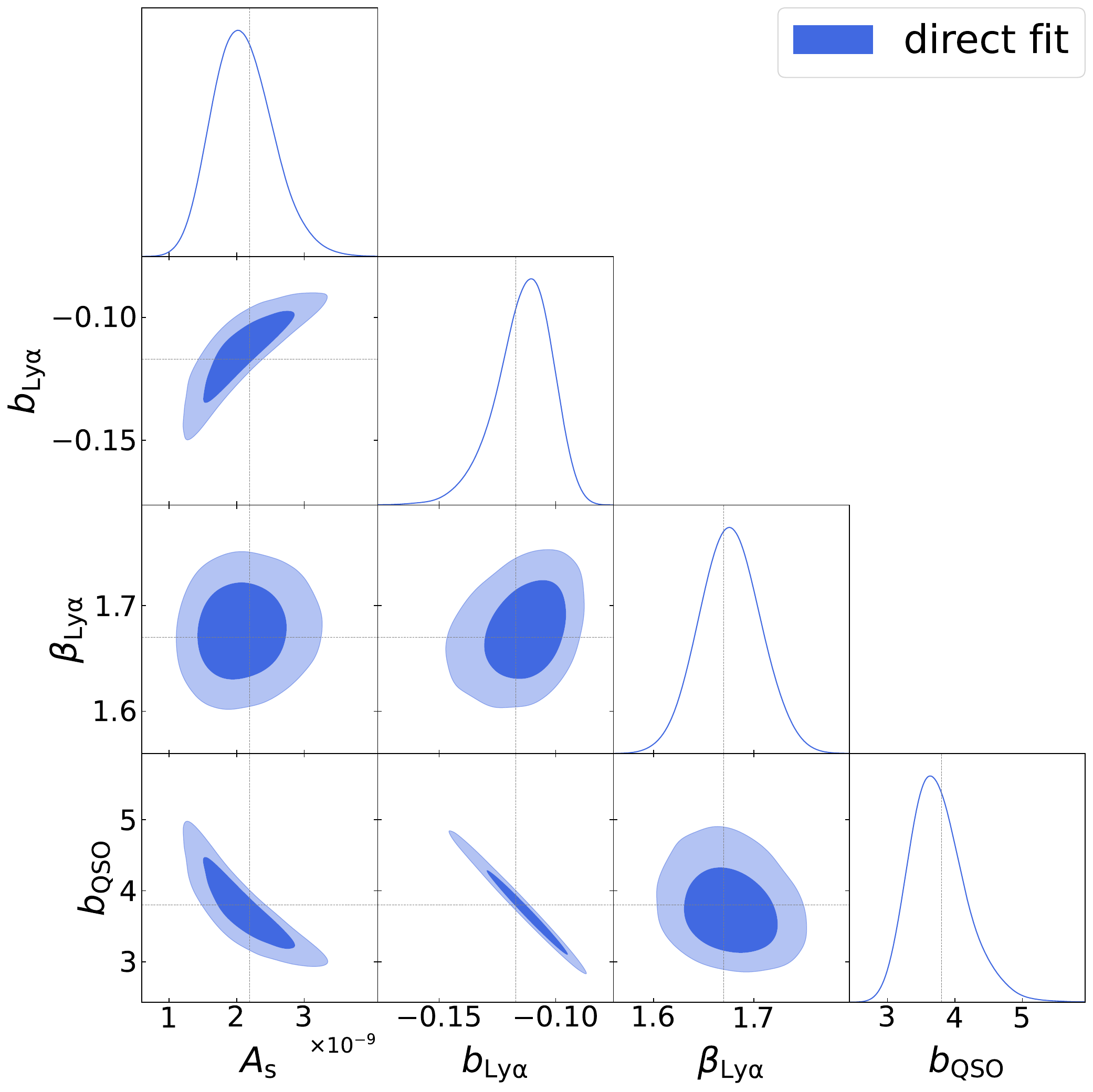}
    \caption{Correlations among the nuisance parameters $\{ b_{\rm{Ly}\alpha},\beta_{\rm{Ly}\alpha},b_{\rm{QSO}} \}$ and $A_{\rm{s}}$, where `direct fit' refers to the inference method described in Sect. \ref{sect:lya_method}. The grey dashed lines mark the fiducial values used to generate the data.}
    \label{fig:growth_degeneracies}
\end{figure}

\section{RESULTS}\label{sect:lya_forecasts}

In this section we present the forecasts produced using the method outlined in Sect. \ref{sect:lya_method} on 3D Ly$\alpha \times$Ly$\alpha$ and Ly$\alpha \times$QSO simplified synthetic correlation functions.
We sample over the cosmological parameters $\boldsymbol{p_{\rm{C}}}=\{H_0, \Omega_{\rm{M}}, \Omega_{\rm{B}} h^2,A_{\rm{s}},n_{\rm{s}}\}$, marginalizing over the astrophysical model parameters $\boldsymbol{p_{\rm{A}}} = \{b_{\rm{Ly}\alpha}, b_{\rm{QSO}}, \beta_{\rm{Ly}\alpha}, \sigma_{\rm{v}}\}$. The fiducial values of these parameters, along with the priors, are listed in Tab.~\ref{tab:full_table}.

In Fig.~\ref{fig:growth_cosmo} we show the results for $\{ H_0, \Omega_{\rm{M}}, \Omega_{\rm{B}} h^2, A_{\rm{s}}, n_{\rm{s}} \}$ using the noiseless mock data vector. In the last column of Tab.~\ref{tab:full_table} we list the one-dimensional marginal constraints for the full set of sampled parameters. 

From Fig.~\ref{fig:growth_cosmo}, it can be seen that we do recover the true values (shown in grey dashed lines) of cosmological parameters $\boldsymbol{p_{\rm{C}}}$ well within $1 \sigma$ (Tab.~\ref{tab:full_table}). The analysis provides a $3.5\%$ constraint on the matter density $\Omega_{\rm{M}}$. 
Clear degeneracies are present between $H_0$ and $\Omega_{\rm{B}} h^{2}$. As previously mentioned, the baryon acoustic oscillation peak measures the product $H_0 r_{\rm{d}}$, which can be expressed as a function of $H_{0}$, $\Omega_{\rm{B}}h^2$ and $\Omega_{\rm{M}}$. Given that we have a good measurement of $\Omega_{\rm{M}}$ from the AP information, the remaining degeneracy is between $ H_0$ and $\Omega_{\rm{B}}h^2$: if there were no other information on either one of them, these two parameters would be fully degenerate. The fact that both $H_0$ and $\Omega_{\rm{B}} h^{2}$ are strongly correlated with the spectral index $n_{\rm{s}}$ could hint that the turnover of the power spectrum is the feature partially breaking the degeneracy. Despite this correlation, we are able to place a constraint on the spectral index, namely $n_s = 0.958^{+0.025}_{-0.035}$. We obtain a $21\%$ constraint on the amplitude of fluctuations $A_{\rm{s}}$, with a corresponding constraint on the amplitude of linear matter fluctuations in spheres of $8~\rm{h^{-1}Mpc}$ of $\sigma_{8}(z_{\rm{eff}}) = 0.317\pm 0.032$.
We will further analyze the constraining power of our analysis against the state-of-the-art results later in Sect. \ref{subsect:lya_cosmo_information}. 

In Fig.~\ref{fig:growth_degeneracies} we show the strong correlation among the linear biases, $b_{\rm{Ly}\alpha}$ and $b_{\rm{QSO}}$, and $A_{\rm{s}}$, which is expected given the functional form of Eqs.~(\ref{eqn:lyalyapower}-\ref{eqn:lyaqsopower}). The Ly$\alpha$ auto-correlation alone would not be able to place a constraint on $A_{\rm{s}}$ since, for $\mu = 0$, we would measure the combination $A_{\rm{s}} b_{\rm{Ly\alpha}}^2$ only, whereas its anisotropy would provide a measurement of $\beta_{\rm{Ly\alpha}}$. On the other hand, the transverse mode of the cross-power spectrum (Eq.~\ref{eqn:lyaqsopower}), combined with the auto-correlation, measures the combination of $A_{\rm{s}} b_{\rm{Ly\alpha}}^2$ and $ b_{\rm{QSO}}/ b_{\rm{Ly\alpha}}$, while through the RSD term we are able to constrain $A_{\rm{s}}$ (or $f\sigma_8$ in compressed analyses).

\section{DISCUSSION}\label{sect:lya_discussion}

In Sect. \ref{subsect:lya_cosmo_information} we provide a direct comparison of the forecasts presented in Sect. \ref{sect:lya_forecasts} and the literature. In particular, we will focus on a comparison with the results on the same synthetic data obtained using the standard BOSS and eBOSS analysis first, as well as the C21 approach. Our goal is to understand whether the compressed analyses successfully capture the cosmological information carried by the 3D Ly$\alpha$ correlation functions and discuss which components of the data are the most informative. In Sect. \ref{subsect:lya_geometry} we discuss how results change when marginalizing over the growth of structure. This is instructive to further understand from where extra information originates.

\subsection{Cosmological information}\label{subsect:lya_cosmo_information}

\begin{figure}
    \centering
    \begin{tikzpicture}
        \node[anchor=south west,inner sep=0] at (0,0) {\includegraphics[scale=0.25]{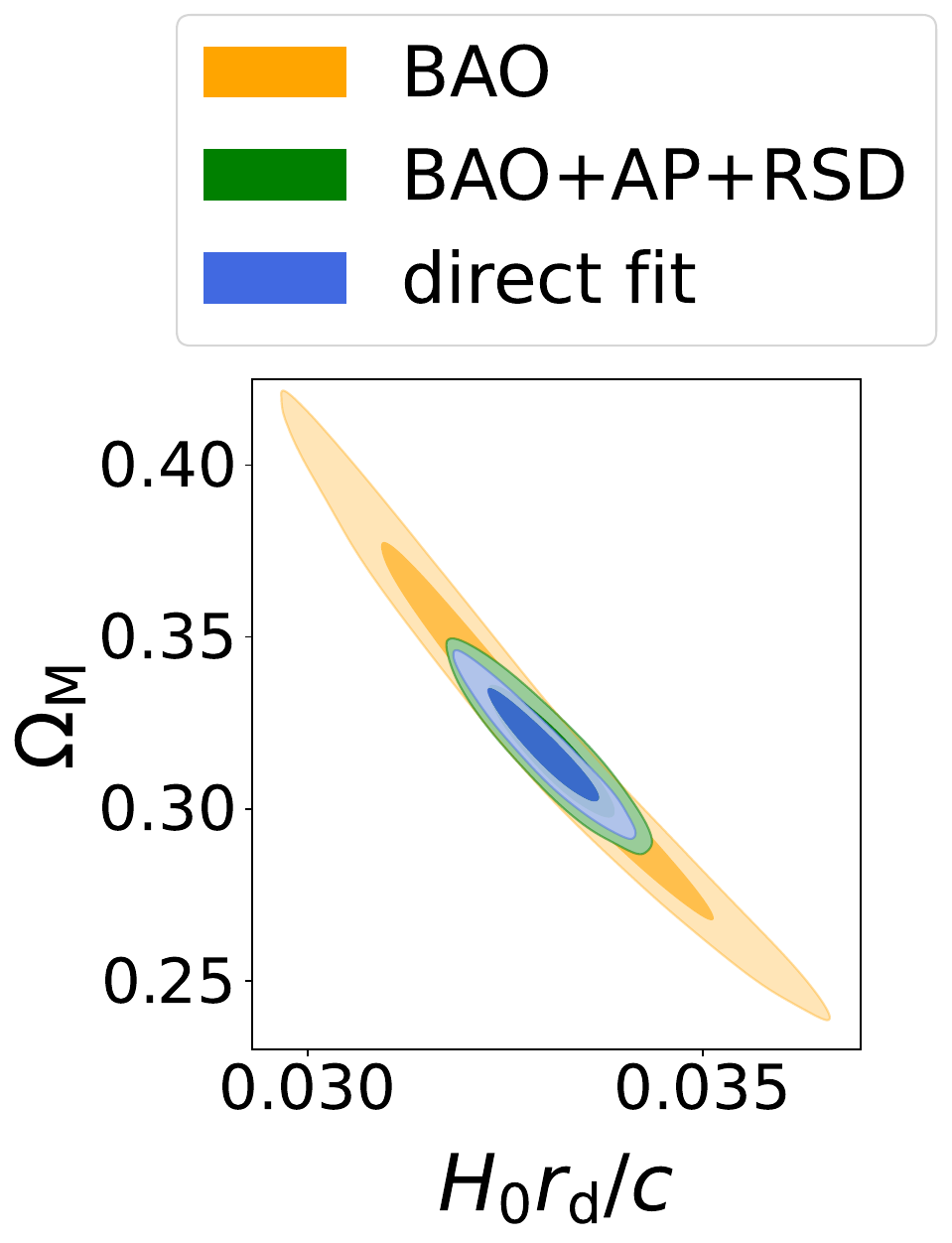}
    \includegraphics[scale=0.305]{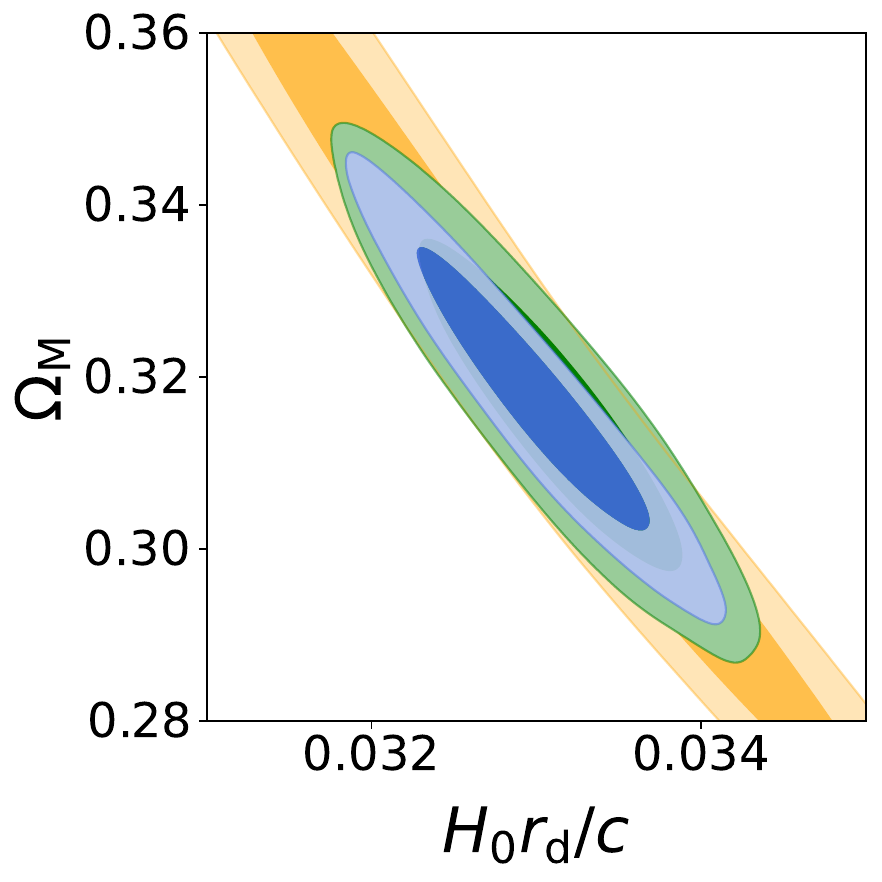}};
        \draw (1.6,2.7) -- (2.95,2.7);
        \draw (1.6,2.7) -- (1.6,1.6);
        \draw (2.95,2.7) -- (2.95,1.6);
        \draw (1.6,1.6) -- (2.95,1.6);
        
        \draw (1.6,2.7) -- (5.2,4.395);
        \draw (1.6,1.6) -- (5.2,0.82);
    \end{tikzpicture}
    \caption{Two-dimensional contour plots of $\{ H_{0}r_{\rm{d}}/c, \Omega_{\rm{M}} \}$, comparing our method (`direct fit') in blue against standard BOSS and eBOSS analysis (`BAO') in orange and C21 (`BAO+AP+RSD') in green. On the right, there is a zoom-in to further highlight the differences among the `direct fit' and `BAO+AP+RSD' methods.} 
    \label{fig:cfr_bao}
\end{figure}
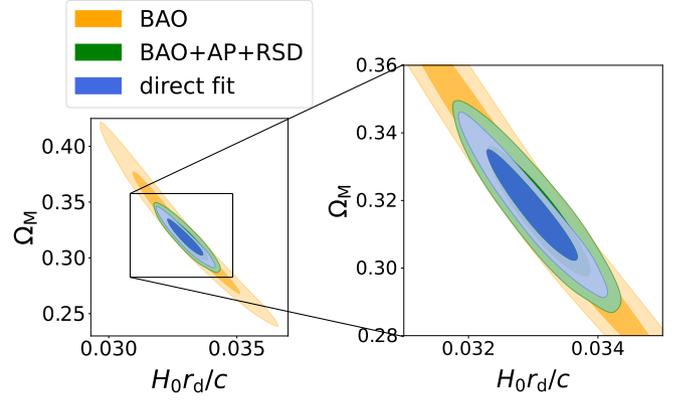

\begin{figure}
    \centering
    \includegraphics[scale=0.29]{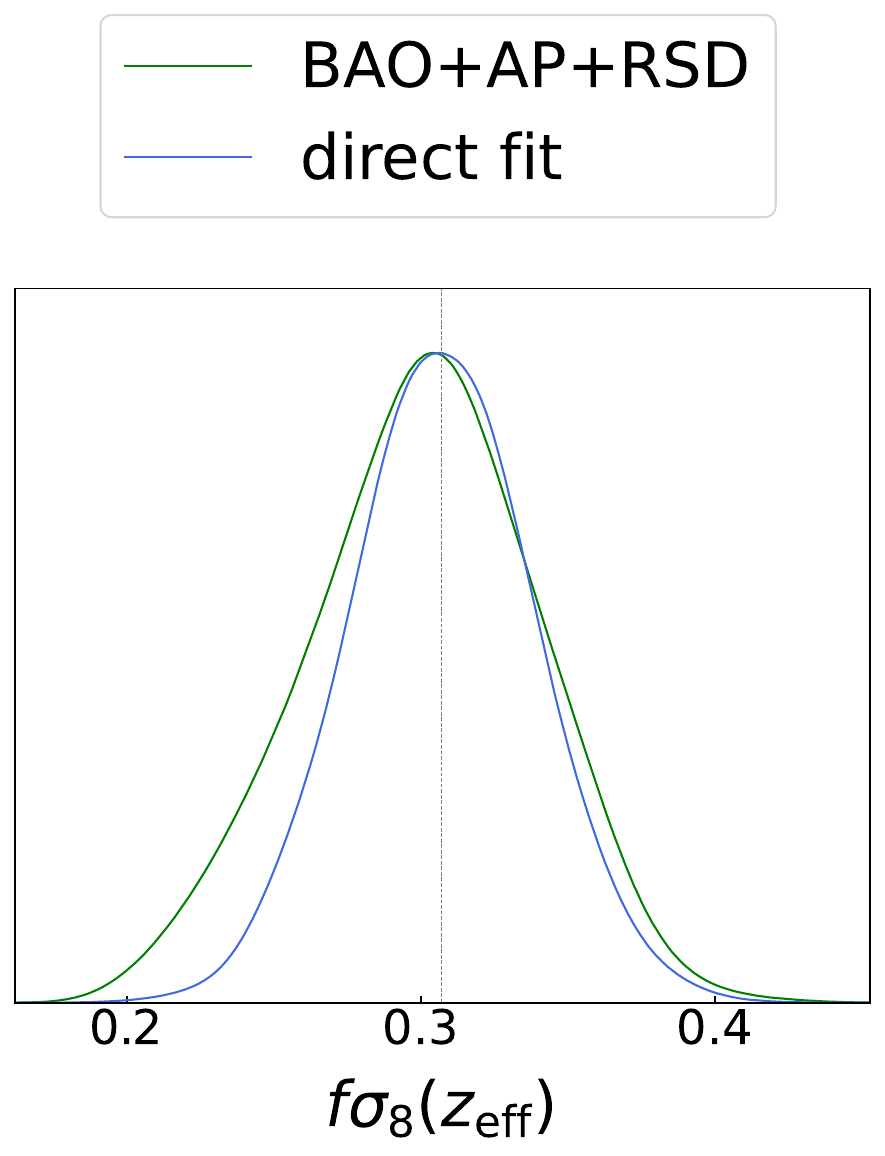}
    \caption{Posterior plot for $f\sigma_{8}(z_{\rm{eff}})$, comparing our method (`direct fit') in blue against C21 (`BAO+AP+RSD') in green.} 
    \label{fig:growth_fs8}
\end{figure}

As mentioned above, in flat $\Lambda$CDM the BAO scale, along and across the line of sight, identifies a banana-shaped degeneracy in the $[H_{0}r_{\rm{d}}, \Omega_{\rm{M}}]$ plane. This justifies that any comparison among methods which have the BAO as a primary feature should necessarily happen in this plane. On the other hand, the Ly$\alpha$-QSO cross-correlation can in principle measure $f \sigma_{8} (z_{\rm{eff}})$ because of the functional form of Eq.~(\ref{eqn:lyaqsopower}). However, because of the degeneracy with the linear biases, the combination with the Ly$\alpha$ auto-correlation is needed. In what follows we will focus on a comparison based on the derived parameters $\boldsymbol{p_{\rm{d}}}=\{ H_0 r_{\rm{d}}/c, f \sigma_{8} (z_{\rm{eff}}) \}$ and $\Omega_{\rm{M}}$. In particular, in Fig.~\ref{fig:cfr_bao} we plot the two-dimensional contours of $\{ H_{0}r_{\rm{d}}/c, \Omega_{\rm{M}} \}$ and in  Fig.~\ref{fig:growth_fs8} the one-dimensional marginal of $ f \sigma_{8} (z_{\rm{eff}})$ for the methods we want to compare. 

\begin{table}
\centering
\begin{tabular}{ l c c c}
\hline
Parameter  & BAO & BAO+AP+RSD &  direct fit\\
\hline
$\Omega_{\rm{M}}$ &  12\% & 4\%  & 3.5 \%\\
$H_{0}r_{\rm{d}}/c $ & 4.5\% & 1.65\%  & 1.43 \%\\
$f \sigma_{8} (z_{\rm{eff}})$ & $-$ & 12.5\%  & 10.4\% 
\end{tabular}
\caption{Constraining power of our method (`direct fit') on the listed parameters, against those from the standard analysis (BAO) and the one of C21 (BAO+AP+RSD).}
\label{tab:der_growth}
\end{table}

As discussed above, standard BOSS and eBOSS  analyses focus on the peak component of the 3D Ly$\alpha$ correlation functions only. For this reason, we will refer to this approach as `BAO' for simplicity. We run this analysis using our noiseless mock data, and the most important result is shown in Fig.~\ref{fig:cfr_bao}. This approach provides $\Omega_{\rm{M}} = 0.32 \pm 0.04$, $H_{0}r_{\rm{d}}/c = 0.0329 \pm 0.0015$, with a constraining power on $H_{0}r_{\rm{d}}/c$ of 4.5\%, a factor of three worse compared to our direct fit (summary in Tab.~\ref{tab:der_growth}). 

Such an improvement was already found by C21, who demonstrated that considering the AP effect from the \textit{smooth} component in addition to the \textit{peak} provides significantly tighter constraints. We present results using their method with the additional RSD information, and we will refer to it as `BAO+AP+RSD'. By running their analysis over our noiseless data, we find that `BAO+AP+RSD' is able to place a constraint on $H_{0}r_{\rm{d}}/c$ of 1.65\%, which is of the same order as for our analysis (Tab.~\ref{tab:der_growth}). As it can be seen in Fig.~\ref{fig:cfr_bao}, $H_{0}r_{\rm{d}}/c$ and $\Omega_{\rm{M}}$ are strongly correlated, with correlation coefficient $0.95$, along the same direction as observed by C21. Our method constrains the growth of structure (Eq.~\ref{eqn:lyalyapower}-\ref{eqn:lyaqsopower}) as well. For this reason, we also compare the RSD information of C21 with ours. The $f \sigma_{8} (z_{\rm{eff}})$ one-dimensional marginals for both methods are plotted in Fig.~\ref{fig:growth_fs8} and the constraining power is given for completion in Tab.~\ref{tab:der_growth}. 
Overall, our method provides tighter constraints, on $H_0 r_{\rm{d}}$, $\Omega_{\rm{M}}$ and $f \sigma_{8} (z_{\rm{eff}})$, with respect to the `BAO+AP+RSD' analysis, by about $16\%$, $17\%$ and $18\%$, respectively.

\begin{figure*}
    \centering
    \includegraphics[scale=0.36]{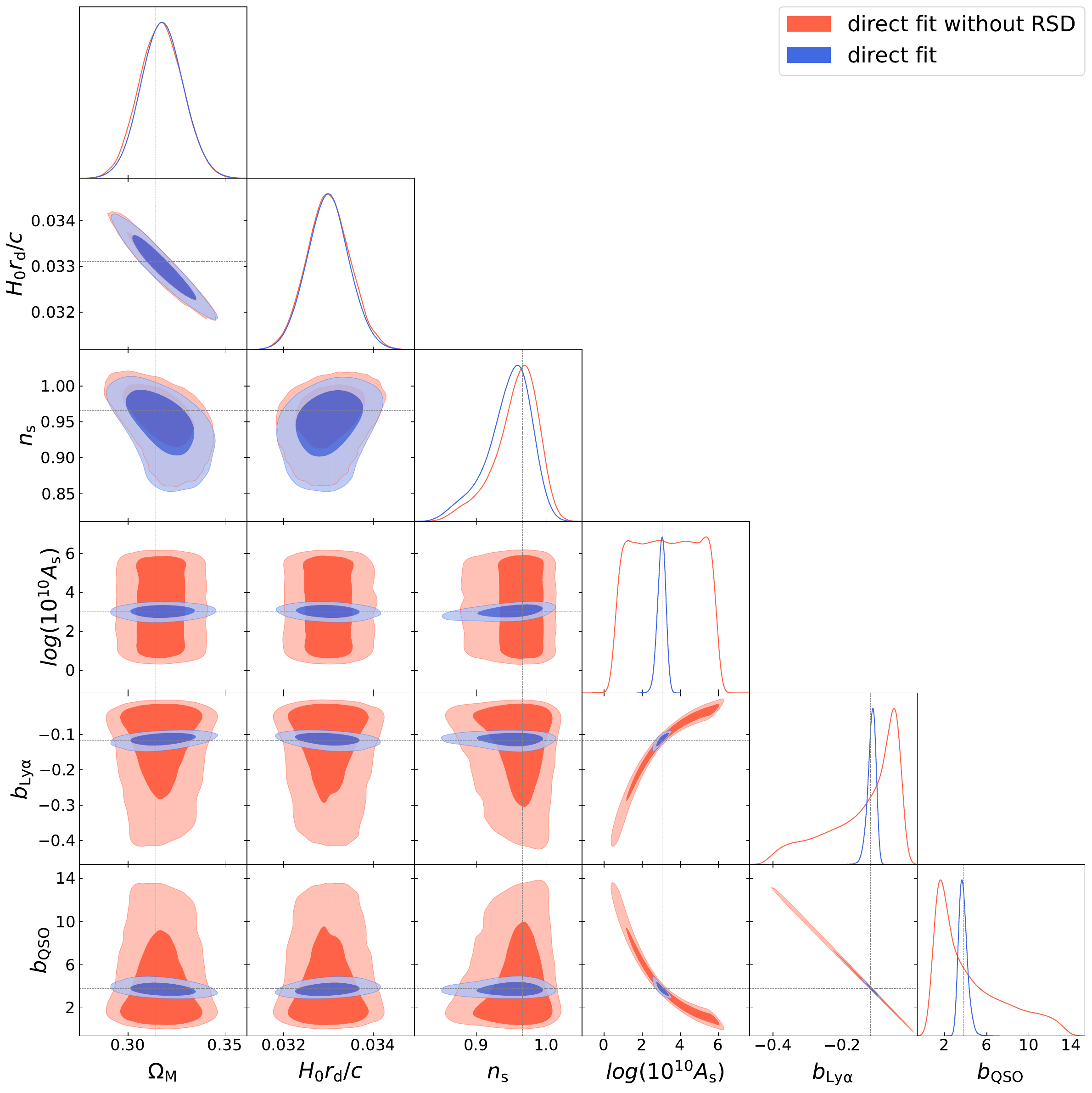}
    \caption{Triangle plot comparing the constraints on $\{ \Omega_{\rm{M}}, H_0 r_{\rm{d}}/c, A_{\rm{s}}, n_{\rm{s}} \}$ and the two linear biases $\{ b_{\rm{Ly}\alpha}, b_{\rm{QSO}} \}$ using our `direct fit' approach in blue against the same fitting method but marginalizing over the growth of structure (`direct fit without RSD') in red. A further explanation of the second approach can be found in Sect. \ref{subsect:lya_geometry}.}
    \label{fig:growthvsgeometry}
\end{figure*}

\subsection{Direct fit analysis without RSD} \label{subsect:lya_geometry}

In order to further investigate where some of the information in the `direct fit' alone is coming from, we repeat the same analysis, marginalizing over the growth of structure. We refer to this case as `direct fit without RSD'.
If we use the quasar RSD parameter defined as $\beta_{\rm{QSO}} = f(z)/b_{\rm{QSO}}$, Eq.~(\ref{eqn:lyaqsopower}) can be rewritten as 

\begin{align}
    P_{\times}(k,\mu_k,z) = & b_{\rm{Ly}\alpha} \left( 1+\beta_{\rm{Ly}\alpha}\mu_k^2 \right)  \\
    & \times b_{\rm{QSO}} \left( 1+\beta_{\rm{QSO}}\mu_k^2 \right)F_{\rm{nl,QSO}}P(k,z) \; , \nonumber
\end{align}

where $\beta_{\rm{QSO}}$ is now an extra nuisance parameter.

The implications of this choice can be directly seen in Fig.~\ref{fig:growthvsgeometry}. In the no-RSD case (red in Figure), the degeneracy among $A_{\rm{s}}$ and the two linear biases extends up to the prior limits, while the information on the other cosmological parameters is preserved. Marginalizing over the growth of structure washes out any constraining power on the linear biases and $A_{\rm{s}}$, while RSD are not constraining any other parameter.

\section{CONCLUSIONS}\label{sect:lya_conclusions}

Given that baryon acoustic oscillations (BAO) produce a distinct feature in 3D Ly$\alpha$ correlation functions, and its properties are well understood, BOSS and eBOSS analyses so far considered only the peak for cosmological inference (`BAO' analysis). A previous analysis conducted by \citet{Cuceu_Font-Ribera_Joachimi_Nadathur_2021} (C21 throughout the paper) highlighted the importance of also considering the broadband component, as it significantly contributes to the overall cosmological information via the Alcock-Paczyński (AP) effect (`BAO+AP' analysis $-$ `BAO+AP+RSD' if redshift space distortions (RSD) are included). Given these premises, in this paper we addressed the question about whether or not the compressed analyses based on BAO, AP and RSD parameters are able to capture all the cosmological information brought by the Ly$\alpha$ correlation functions. We performed a full shape analysis without any of the above parameters and instead directly inferred cosmology (`direct fit' analysis).

The inference framework we used is \textsc{cobaya}, for which we implemented an \textit{ad-hoc} gaussian likelihood based on the \textsc{vega} package, as extensively described in Sect. \ref{sect:lya_method}. 

We performed `BAO', `BAO+AP+RSD' and `direct fit' analyses on the same set of synthetic Ly$\alpha \times$Ly$\alpha$ auto- and Ly$\alpha \times$QSO cross-correlations, which include distortion effects due to continuum fitting, but no other contaminants, and ran the inference over $\boldsymbol{p_{\rm{C}}}=\{H_0, \Omega_{\rm{M}}, \Omega_{\rm{B}} h^2,A_{\rm{s}},n_{\rm{s}}\}$. We also marginalized over the nuisance astrophysical parameters $\boldsymbol{p_{\rm{A}}} = \{b_{\rm{Ly}\alpha}, b_{\rm{QSO}}, \beta_{\rm{Ly}\alpha}, \sigma_{\rm{v}}\}$, which are the Ly$\alpha$ and quasars linear biases, the Ly$\alpha$ RSD term and the velocity dispersion of quasars respectively.

We were able to measure the matter density parameter $\Omega_{\rm{M}}$ and the amplitude of primordial fluctuations $A_{\rm{s}}$ with a precision of $3.5\%$ and $21\%$, respectively, and $f \sigma_{8} (z_{\rm{eff}})$ at $10.4\%$, which is a noteworthy result given the few measurements of this parameter at $z>2$. For these parameters we do not obtain a significant improvement in constraining power with respect to the `BAO+AP+RSD' approach. However, we obtain a constraint of the spectral index $n_{\rm{s}}$ at the $\sim 3-4 \%$ level. Similarly to the findings and solutions put forward in \citet{Brieden_2021}, we could account for this extra information by adding a slope parameter to the compressed analysis.

The robustness of the `direct fit' method against systematics must be tested. In forthcoming work, it would be interesting to investigate whether the compressed analysis is more robust, given it measures specific physical effects that are well understood. A further natural next step should be including contaminants to the analysis. The constraining power in the spectral index that we achieve could be affected in particular by Damped Ly$\alpha$ System (DLA) contamination \citep{McQuinn_2011, Font_Ribera_2012} and fluctuations in the UV background \citep{Pontzen_2014, Gontcho_A_Gontcho_2014}, which would change the correlation function in a way that could mimic $n_{\rm{s}}$. Further improvements to the analysis could come from varying $r_{\rm{min}}$ and $r_{\rm{max}}$, and also checking how effects of continuum distortion consequently behave.

We conclude by recalling that soon DESI will provide even better measurements of Ly$\alpha$ correlations. Therefore this kind of study is key to finding the optimal approach to infer cosmology from the data.

\section*{Acknowledgements}

This work was partially enabled by funding from the UCL Cosmoparticle Initiative. AC acknowledges support from the United States Department of Energy, Office of High Energy Physics under Award Number DE-SC-0011726. AFR acknowledges support by the program Ramon y Cajal (RYC-2018-025210) of the Spanish Ministry of Science and Innovation and from the European Union’s Horizon Europe research and innovation programme (COSMO-LYA, grant agreement 101044612).
IFAE is partially funded by the CERCA program of the Generalitat de Catalunya. BJ acknowledges support by STFC Consolidated Grant ST/V000780/1. PL acknowledges STFC Consolidated Grants ST/R000476/1 and ST/T000473/1.
For the purpose of open access, the authors have applied a creative commons attribution (CC BY) licence to any author-accepted manuscript version arising.

\section*{Data Availability}

The code is publicly available at \url{https://github.com/frgerardi/LyA_directfit}. The data underlying this article will be shared on reasonable request to the corresponding author.
 



\bibliographystyle{mnras}
\bibliography{biblio} 

\begin{thebibliography}{}
\makeatletter
\relax
\def\mn@urlcharsother{\let\do\@makeother \do\$\do\&\do\#\do\^\do\_\do\%\do\~}
\def\mn@doi{\begingroup\mn@urlcharsother \@ifnextchar [ {\mn@doi@}
  {\mn@doi@[]}}
\def\mn@doi@[#1]#2{\def\@tempa{#1}\ifx\@tempa\@empty \href
  {http://dx.doi.org/#2} {doi:#2}\else \href {http://dx.doi.org/#2} {#1}\fi
  \endgroup}
\def\mn@eprint#1#2{\mn@eprint@#1:#2::\@nil}
\def\mn@eprint@arXiv#1{\href {http://arxiv.org/abs/#1} {{\tt arXiv:#1}}}
\def\mn@eprint@dblp#1{\href {http://dblp.uni-trier.de/rec/bibtex/#1.xml}
  {dblp:#1}}
\def\mn@eprint@#1:#2:#3:#4\@nil{\def\@tempa {#1}\def\@tempb {#2}\def\@tempc
  {#3}\ifx \@tempc \@empty \let \@tempc \@tempb \let \@tempb \@tempa \fi \ifx
  \@tempb \@empty \def\@tempb {arXiv}\fi \@ifundefined
  {mn@eprint@\@tempb}{\@tempb:\@tempc}{\expandafter \expandafter \csname
  mn@eprint@\@tempb\endcsname \expandafter{\@tempc}}}

\bibitem[\protect\citeauthoryear{{Alam} et~al.,}{{Alam}
  et~al.}{2017}]{2017MNRAS.470.2617A}
{Alam} S.,  et~al., 2017, \mn@doi [\mnras] {10.1093/mnras/stx721}, \href
  {https://ui.adsabs.harvard.edu/abs/2017MNRAS.470.2617A} {470, 2617}

\bibitem[\protect\citeauthoryear{{Alam} et~al.,}{{Alam}
  et~al.}{2021}]{2021PhRvD.103h3533A}
{Alam} S.,  et~al., 2021, \mn@doi [\prd] {10.1103/PhysRevD.103.083533}, \href
  {https://ui.adsabs.harvard.edu/abs/2021PhRvD.103h3533A} {103, 083533}

\bibitem[\protect\citeauthoryear{{Alcock} \& {Paczynski}}{{Alcock} \&
  {Paczynski}}{1979}]{1979Natur.281..358A}
{Alcock} C.,  {Paczynski} B.,  1979, \mn@doi [\nat] {10.1038/281358a0}, \href
  {https://ui.adsabs.harvard.edu/abs/1979Natur.281..358A} {281, 358}

\bibitem[\protect\citeauthoryear{Arinyo-i Prats, Miralda-Escud\'e, Viel  \&
  Cen}{Arinyo-i Prats et~al.}{2015}]{Arinyo-i-Prats:2015vqa}
Arinyo-i Prats A.,  Miralda-Escud\'e J.,  Viel M.,   Cen R.,  2015, \mn@doi
  [JCAP] {10.1088/1475-7516/2015/12/017}, 12, 017

\bibitem[\protect\citeauthoryear{Aubourg et~al.,}{Aubourg
  et~al.}{2015}]{PhysRevD.92.123516}
Aubourg E.,  et~al., 2015, \mn@doi [Phys. Rev. D] {10.1103/PhysRevD.92.123516},
  92, 123516

\bibitem[\protect\citeauthoryear{Bautista et~al.}{Bautista
  et~al.}{2017}]{Bautista:2017zgn}
Bautista J.~E.,  et~al., 2017, \mn@doi [Astron. Astrophys.]
  {10.1051/0004-6361/201730533}, 603, A12

\bibitem[\protect\citeauthoryear{Brieden, Gil-Mar{\'{\i}}n  \& Verde}{Brieden
  et~al.}{2021}]{Brieden_2021}
Brieden S.,  Gil-Mar{\'{\i}}n H.,   Verde L.,  2021, \mn@doi [Journal of
  Cosmology and Astroparticle Physics] {10.1088/1475-7516/2021/12/054}, 2021,
  054

\bibitem[\protect\citeauthoryear{Busca et~al.,}{Busca
  et~al.}{2013}]{Busca_Delubac_Rich_Bailey_Font-Ribera_Kirkby_Le_Goff_Pieri_Slosar_Aubourg_et_al._2013}
Busca N.~G.,  et~al., 2013, \mn@doi [Astronomy & Astrophysics]
  {10.1051/0004-6361/201220724}, 552, A96

\bibitem[\protect\citeauthoryear{Cole et~al.}{Cole
  et~al.}{2005}]{2dFGRS:2005yhx}
Cole S.,  et~al., 2005, \mn@doi [Mon. Not. Roy. Astron. Soc.]
  {10.1111/j.1365-2966.2005.09318.x}, 362, 505

\bibitem[\protect\citeauthoryear{Cuceu, Font-Ribera  \& Joachimi}{Cuceu
  et~al.}{2020}]{Cuceu_2020}
Cuceu A.,  Font-Ribera A.,   Joachimi B.,  2020, \mn@doi [Journal of Cosmology
  and Astroparticle Physics] {10.1088/1475-7516/2020/07/035}, 2020, 035

\bibitem[\protect\citeauthoryear{Cuceu, Font-Ribera, Joachimi  \&
  Nadathur}{Cuceu et~al.}{2021}]{Cuceu_Font-Ribera_Joachimi_Nadathur_2021}
Cuceu A.,  Font-Ribera A.,  Joachimi B.,   Nadathur S.,  2021, \mn@doi [Monthly
  Notices of the Royal Astronomical Society] {10.1093/mnras/stab1999}, 506,
  5439–5450

\bibitem[\protect\citeauthoryear{{DESI Collaboration} et~al.,}{{DESI
  Collaboration} et~al.}{2016}]{https://doi.org/10.48550/arxiv.1611.00036}
{DESI Collaboration} et~al., 2016, The DESI Experiment Part I:
  Science,Targeting, and Survey Design, \mn@doi{10.48550/ARXIV.1611.00036},
  \url {https://arxiv.org/abs/1611.00036}

\bibitem[\protect\citeauthoryear{Eisenstein et~al.}{Eisenstein
  et~al.}{2005}]{SDSS:2005xqv}
Eisenstein D.~J.,  et~al., 2005, \mn@doi [Astrophys. J.] {10.1086/466512}, 633,
  560

\bibitem[\protect\citeauthoryear{Farr et~al.,}{Farr et~al.}{2020}]{Farr_2020}
Farr J.,  et~al., 2020, \mn@doi [Journal of Cosmology and Astroparticle
  Physics] {10.1088/1475-7516/2020/03/068}, 2020, 068

\bibitem[\protect\citeauthoryear{Font-Ribera et~al.,}{Font-Ribera
  et~al.}{2012}]{Font_Ribera_2012}
Font-Ribera A.,  et~al., 2012, \mn@doi [Journal of Cosmology and Astroparticle
  Physics] {10.1088/1475-7516/2012/11/059}, 2012, 059

\bibitem[\protect\citeauthoryear{Font-Ribera et~al.}{Font-Ribera
  et~al.}{2014}]{BOSS:2013igd}
Font-Ribera A.,  et~al., 2014, \mn@doi [JCAP] {10.1088/1475-7516/2014/05/027},
  05, 027

\bibitem[\protect\citeauthoryear{Gontcho, Miralda-Escud{\'{e} }  \&
  Busca}{Gontcho et~al.}{2014}]{Gontcho_A_Gontcho_2014}
Gontcho S. G.~A.,  Miralda-Escud{\'{e} } J.,   Busca N.~G.,  2014, \mn@doi
  [Monthly Notices of the Royal Astronomical Society] {10.1093/mnras/stu860},
  442, 187

\bibitem[\protect\citeauthoryear{Hamilton}{Hamilton}{2000}]{10.1046/j.1365-8711.2000.03071.x}
Hamilton A. J.~S.,  2000, \mn@doi [Monthly Notices of the Royal Astronomical
  Society] {10.1046/j.1365-8711.2000.03071.x}, 312, 257

\bibitem[\protect\citeauthoryear{Howlett, Lewis, Hall  \& Challinor}{Howlett
  et~al.}{2012}]{Howlett_2012}
Howlett C.,  Lewis A.,  Hall A.,   Challinor A.,  2012, \mn@doi [Journal of
  Cosmology and Astroparticle Physics] {10.1088/1475-7516/2012/04/027}, 2012,
  027

\bibitem[\protect\citeauthoryear{Ivanov, Simonovi{\'{c} }  \&
  Zaldarriaga}{Ivanov et~al.}{2020}]{Ivanov_2020}
Ivanov M.~M.,  Simonovi{\'{c} } M.,   Zaldarriaga M.,  2020, \mn@doi [Journal
  of Cosmology and Astroparticle Physics] {10.1088/1475-7516/2020/05/042},
  2020, 042

\bibitem[\protect\citeauthoryear{Kirkby et~al.}{Kirkby
  et~al.}{2013}]{Kirkby:2013fh}
Kirkby D.,  et~al., 2013, \mn@doi [JCAP] {10.1088/1475-7516/2013/03/024}, 03,
  024

\bibitem[\protect\citeauthoryear{Lewis}{Lewis}{2013}]{PhysRevD.87.103529}
Lewis A.,  2013, \mn@doi [Phys. Rev. D] {10.1103/PhysRevD.87.103529}, 87,
  103529

\bibitem[\protect\citeauthoryear{Lewis \& Bridle}{Lewis \&
  Bridle}{2002}]{PhysRevD.66.103511}
Lewis A.,  Bridle S.,  2002, \mn@doi [Phys. Rev. D]
  {10.1103/PhysRevD.66.103511}, 66, 103511

\bibitem[\protect\citeauthoryear{Lewis, Challinor  \& Lasenby}{Lewis
  et~al.}{2000}]{Lewis_2000}
Lewis A.,  Challinor A.,   Lasenby A.,  2000, \mn@doi [The Astrophysical
  Journal] {10.1086/309179}, 538, 473

\bibitem[\protect\citeauthoryear{McDonald \& Eisenstein}{McDonald \&
  Eisenstein}{2007}]{PhysRevD.76.063009}
McDonald P.,  Eisenstein D.~J.,  2007, \mn@doi [Phys. Rev. D]
  {10.1103/PhysRevD.76.063009}, 76, 063009

\bibitem[\protect\citeauthoryear{McQuinn \& White}{McQuinn \&
  White}{2011}]{McQuinn_2011}
McQuinn M.,  White M.,  2011, \mn@doi [Monthly Notices of the Royal
  Astronomical Society] {10.1111/j.1365-2966.2011.18855.x}, 415, 2257

\bibitem[\protect\citeauthoryear{Neal}{Neal}{2005}]{https://doi.org/10.48550/arxiv.math/0502099}
Neal R.~M.,  2005, Taking Bigger Metropolis Steps by Dragging Fast Variables,
  \mn@doi{10.48550/ARXIV.MATH/0502099}, \url
  {https://arxiv.org/abs/math/0502099}

\bibitem[\protect\citeauthoryear{Perlmutter et~al.,}{Perlmutter
  et~al.}{1999}]{Perlmutter_1999}
Perlmutter S.,  et~al., 1999, \mn@doi [The Astrophysical Journal]
  {10.1086/307221}, 517, 565

\bibitem[\protect\citeauthoryear{Pontzen \& Governato}{Pontzen \&
  Governato}{2014}]{Pontzen_2014}
Pontzen A.,  Governato F.,  2014, \mn@doi [Nature] {10.1038/nature12953}, 506,
  171

\bibitem[\protect\citeauthoryear{Ram{\'{\i}}rez-P{\'{e}}rez, Sanchez, Alonso
  \& Font-Ribera}{Ram{\'{\i}}rez-P{\'{e}}rez et~al.}{2022}]{Ram_rez_P_rez_2022}
Ram{\'{\i}}rez-P{\'{e}}rez C.,  Sanchez J.,  Alonso D.,   Font-Ribera A.,
  2022, \mn@doi [Journal of Cosmology and Astroparticle Physics]
  {10.1088/1475-7516/2022/05/002}, 2022, 002

\bibitem[\protect\citeauthoryear{Riess et~al.,}{Riess
  et~al.}{1998}]{Riess_1998}
Riess A.~G.,  et~al., 1998, \mn@doi [The Astronomical Journal]
  {10.1086/300499}, 116, 1009

\bibitem[\protect\citeauthoryear{Seljak}{Seljak}{2012}]{Seljak_2012}
Seljak U.,  2012, \mn@doi [Journal of Cosmology and Astroparticle Physics]
  {10.1088/1475-7516/2012/03/004}, 2012, 004

\bibitem[\protect\citeauthoryear{{Seo} \& {Eisenstein}}{{Seo} \&
  {Eisenstein}}{2003}]{2003ApJ...598..720S}
{Seo} H.-J.,  {Eisenstein} D.~J.,  2003, \mn@doi [\apj] {10.1086/379122}, \href
  {https://ui.adsabs.harvard.edu/abs/2003ApJ...598..720S} {598, 720}

\bibitem[\protect\citeauthoryear{Slosar et~al.}{Slosar
  et~al.}{2013}]{Slosar:2013fi}
Slosar A.,  et~al., 2013, \mn@doi [JCAP] {10.1088/1475-7516/2013/04/026}, 04,
  026

\bibitem[\protect\citeauthoryear{{Torrado} \& {Lewis}}{{Torrado} \&
  {Lewis}}{2019}]{2019ascl.soft10019T}
{Torrado} J.,  {Lewis} A.,  2019, {Cobaya: Bayesian analysis in cosmology},
  Astrophysics Source Code Library, record ascl:1910.019 (\mn@eprint {ascl}
  {1910.019})

\bibitem[\protect\citeauthoryear{Torrado \& Lewis}{Torrado \&
  Lewis}{2021}]{Torrado_2021}
Torrado J.,  Lewis A.,  2021, \mn@doi [Journal of Cosmology and Astroparticle
  Physics] {10.1088/1475-7516/2021/05/057}, 2021, 057

\bibitem[\protect\citeauthoryear{{Tr{\"o}ster} et~al.,}{{Tr{\"o}ster}
  et~al.}{2020}]{troester20}
{Tr{\"o}ster} T.,  et~al., 2020, \mn@doi [\aap] {10.1051/0004-6361/201936772},
  \href {https://ui.adsabs.harvard.edu/abs/2020A&A...633L..10T} {633, L10}

\bibitem[\protect\citeauthoryear{Youles et~al.,}{Youles
  et~al.}{2022}]{https://doi.org/10.48550/arxiv.2205.06648}
Youles S.,  et~al., 2022, The effect of quasar redshift errors on
  Lyman-$\alpha$ forest correlation functions,
  \mn@doi{10.48550/ARXIV.2205.06648}, \url {https://arxiv.org/abs/2205.06648}

\bibitem[\protect\citeauthoryear{d{\textquotesingle}Amico, Gleyzes, Kokron,
  Markovic, Senatore, Zhang, Beutler  \&
  Gil-Mar{\'{\i}}n}{d{\textquotesingle}Amico et~al.}{2020}]{d_Amico_2020}
d{\textquotesingle}Amico G.,  Gleyzes J.,  Kokron N.,  Markovic K.,  Senatore
  L.,  Zhang P.,  Beutler F.,   Gil-Mar{\'{\i}}n H.,  2020, \mn@doi [Journal of
  Cosmology and Astroparticle Physics] {10.1088/1475-7516/2020/05/005}, 2020,
  005

\bibitem[\protect\citeauthoryear{{du Mas des Bourboux} et~al.,}{{du Mas des
  Bourboux} et~al.}{2017}]{dMdBourboux_2017}
{du Mas des Bourboux} H.,  et~al., 2017, \mn@doi [\aap]
  {10.1051/0004-6361/201731731}, \href
  {https://ui.adsabs.harvard.edu/abs/2017A&A...608A.130D} {608, A130}

\bibitem[\protect\citeauthoryear{du~Mas~des Bourboux et~al.,}{du~Mas~des
  Bourboux
  et~al.}{2020}]{Bourboux_Rich_Font-Ribera_Agathe_Farr_Etourneau_Goff_Cuceu_Balland_Bautista_et_al._2020}
du~Mas~des Bourboux H.,  et~al., 2020, \mn@doi [The Astrophysical Journal]
  {10.3847/1538-4357/abb085}, 901, 153

\makeatother
\end{thebibliography}





\bsp	
\label{lastpage}
\end{document}